# Large Nonreciprocity of Shear-Horizontal Surface Acoustic Waves induced by Magnetoelastic Bilayers

Mingxian Huang[1,&], Yuanyuan Liu[1,&], Wenbin Hu[1], Yutong Wu[2], Wen Wang[2], Wei He[3], Huaiwu Zhang[1] and Feiming Bai[1,*]

[1]School of Electronic Science and Engineering, University of Electronic Science and Technology of China, Chengdu, 610054, China

[2]Institute of Acoustics, Chinese Academy of Sciences, Beijing, 100190, China

[3]State Key Laboratory of Magnetism and Beijing National Laboratory for Condensed Matter Physics, Institute of Physics, Chinese Academy of Sciences, Beijing, 100190, China.

& These authors are co-first authors of the article.

* To whom correspondence should be addressed. Electronic mails:

fmbai@uestc.edu.cn

**Abstract**

We report large nonreciprocity in the transmission of shear-horizontal surface acoustic waves (SAWs) on LiTaO$_3$ substrate coated with a FeCoSiB/NiFeCu magnetoelastic bilayer. The large difference in saturation magnetization of the two layers not only brings nonreciprocal spin waves (SWs), but also ensures the phonon-magnon (SAWs-SWs) coupling at relatively low wavenumbers. It is found that the angle between the magnetization and the wavevector play important roles in determining the strength of magnetoelastic coupling and nonreciprocity, simultaneously. A large nonreciprocal transmission of SAWs about 30 dB (i.e. 60 dB/mm) is demonstrated at 2.33 GHz. In addition, the dispersion relation between coupled SH-SAWs and nonreciprocal SWs is developed, which provide a good insight into the observed phenomena. Our results offer a convenient approach to implement nonreciprocal SAW isolators or circulators.

# I. Introduction

Surface acoustic waves (SAWs) are essential for information and communications technology due to its short wavelength and low propagation loss [1-3]. The propagation of SAWs is generally reciprocal. However, it has been found recently that nonreciprocal propagation of SAWs can be achieved by breaking the space-time inversion symmetry via introducing magnetoelastic coupling (MEC) [4-17]. When both the resonance frequency and wavevector of SAWs match those of spin waves (SWs), intense phonon-magnon coupling occurs with significant energy absorption due to magnetization precession, meanwhile, the propagation loss of SAWs remains low when SAWs and SWs are uncoupled. This unique feature offers the magnetoacoustic devices with high isolation and low insertion loss, which greatly promotes the development of nonreciprocal RF/microwave isolator or circulator devices.

At present, the means of achieving nonreciprocal propagation of SAWs via MEC can be divided into two types. One is to use the chirality of the driven field excited by SAWs under MEC [4-8], the other is to use special magnetic structures with nonreciprocal SWs dispersion relationship [9-17]. Since the chirality of the MEC driven field is restricted by the type of SAWs, different magnetic structures with nonreciprocal SW dispersion relationship have been paid much attention [18-24], which also provide a great degree of freedom for device design. Recently, M. Kuß et al. reported nonreciprocal transmission of SAWs in CoFeB/Pt and CoFeB/Au/NiFe structures using Dzyaloshinskii–Moriya interaction (DMI) [9] and interlayer magnetic dipolar coupling [10], respectively. Later, Piyush J. Shah et al. reported nonreciprocity of SAWs using synthetic antiferromagnet (SAFM) of $FeGaB/Al_2O_3/FeGaB$ [13].

Since strong SWs nonreciprocity tends to appear at large wavenumbers ($k$), the wavelengths ($\lambda$)

of SAWs were designed very short to match that of SWs. Previously, most SAWs devices typically achieve strong nonreciprocity at 5-8 GHz [9,10,16,17], which is much higher than commercial SAWs devices, and also put harsh requirements on nanolithography. This difficulty can be mitigated by either selecting piezoelectric substrate and SAW modes with high phase velocity [25] or reducing the spin wave resonance (SWR) frequency. Magnetoelastic bilayers allow a wide range of tuning of ferromagnetic resonance (FMR) frequencies by introducing a low saturation magnetization layer [26-28]. In addition, different from SAFM, which requires precise control of the thickness of the space layer over a large area, the preparation of the magnetoelastic bilayer is much easier. Of particular interests is that the graded saturation magnetization of bilayered or multilayered structure can break the space symmetry in the thickness direction and offer SWs nonreciprocity [24,29]. Therefore, upon coating such magnetoelastic bilayers on SAWs devices, large nonreciprocal transmission of SAWs is expected at relatively low $k$ values.

In current work, we selected NiFeCu with a low saturation magnetization of 0.56 Tesla and magnetostrictive FeCoSiB with a high saturation magnetization of 1.55 Tesla to form magnetoelastic bilayers. Both layers feature with very low Gilbert damping factors [30]. The spin-wave nonreciprocity of FeCoSiB/NiFeCu bilayers were simulated upon varying the total thickness, thickness ratio and the angle between the magnetization and the spin wave vector. In addition, the MEC between nonreciprocal SWs and SH-SAWs excited on a $LiTaO_3$ substrate was also taken into consideration during the design of magnetic bilayer structure. Our fabricated SH-SAWs delay lines feature giant nonreciprocity up to ~30 dB (i.e. 60 dB/mm) at 2.33 GHz.

## II. Theory

**A. Spin-wave nonreciprocity in FeCoSiB/NiFeCu bilayers**

Figure 1(a) shows the magnetic system setup in our simulation, where the spin wave propagates along the x-axis, and the z-axis is the normal direction of the FeCoSiB/NiFeCu bilayer. $\varphi$ represents the angle between the magnetization $M$ and the spin wave vector $k$, and can be varied by setting an in-plane uniaxial anisotropic field along different directions. The magnetic dynamics of 130 μm×130 μm magnetic bilayers are solved by using periodic boundary conditions. Upon applying a perturbed magnetic field $\mathbf{h} = \frac{\sin(2\pi f_0 t)}{2\pi f_0 t}\hat{z}$ at the plane of x=0, SWs are excited in sinc-pulse in space with the cut-off frequency of $f_0 = 10\,\text{GHz}$.

A micromagnetic simulation program (MuMax3) was used to solve the spin-wave dispersion relation of the magnetic system above. The dynamics of the unit magnetization $\vec{\mathbf{m}}(\vec{r},t)$ can be described by the Landau–Lifshitz–Gilbert (LLG) equation [31,32]:

$$\frac{\partial \vec{\mathbf{m}}(\vec{r},t)}{\partial t} = \gamma \frac{1}{1+\alpha^2}\left(\vec{\mathbf{m}}\times\vec{\mathbf{B}}_{\text{eff}} + \alpha\left(\vec{\mathbf{m}}\times\left(\vec{\mathbf{m}}\times\vec{\mathbf{B}}_{\text{eff}}\right)\right)\right) \quad (1)$$

where $\gamma$ is the gyromagnetic ratio, $\alpha$ is the Gilbert damping factor and $\vec{\mathbf{B}}_{\text{eff}}$ represents the effective field. The effective magnetic field $\vec{\mathbf{B}}_{\text{eff}}$ includes the contributions of the dipole field $\vec{\mathbf{B}}_{\text{d}}$, the Heisenberg exchange field $\vec{\mathbf{B}}_{\text{exch}}$ and the anisotropic field $\vec{\mathbf{B}}_{\text{anis}}$. The dipole field $\vec{\mathbf{B}}_{\text{d}}$ can be expressed as:

$$\vec{\mathbf{B}}_{\text{d}} = \int_V \hat{\mathbf{K}}*\vec{\mathbf{M}}dV \quad (2)$$

where $\hat{\mathbf{K}}$ is the demagnetizing kernel [23,24,29,33,34] and $\vec{\mathbf{M}} = M_s\vec{\mathbf{m}}$ is the unnormalized magnetization with $M_s$ the saturation magnetization. The Heisenberg exchange field $\vec{\mathbf{B}}_{\text{exch}}$ can be given by [33,35,36,37]:

$$\vec{B}_{exch} = 2\frac{A}{M_s}\Delta\vec{m} \qquad (3)$$

and $A$ is the magnetic exchange stiffness. The FeCoSiB layer and the NiFeCu layer are connected by dipole and exchange interactions, and the exchange field at the interface $\vec{B}_{exch}^{int}$ can be expressed as [33]:

$$\vec{B}_{exch}^{int} = 2S\frac{2\frac{A_1}{M_s^A}\frac{A_2}{M_s^B}}{\frac{A_1}{M_s^A}+\frac{A_2}{M_s^B}}\sum_i \frac{(\vec{m}_i-\vec{m})}{\Delta_i^2} \qquad (4)$$

where $S$ is a scaling factor for interface coupling, and was set to be 1. $\Delta_i$ is the cell size (2.00×2.00×2.00 nm$^3$) in the direction of neighboring $i$. Here, we set $M_s^A = 1.55$ T, $A_1 = 2.5\times10^{-11}$ J/m for FeCoSiB and $M_s^B = 0.56$ T, $A_2 = 0.35\times10^{-11}$ J/m for NiFeCu [27].

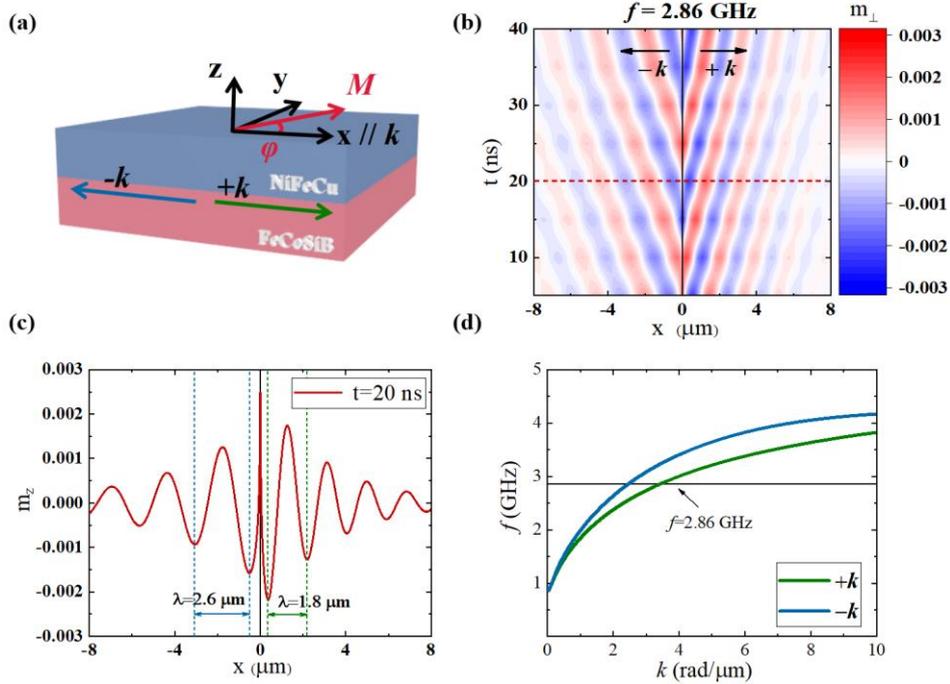

FIG. 1. (a) The magnetic bilayers system setup in the simulation. (b) The spatial and temporal distribution of $m_z$ for the FeCoSiB (30nm) /NiFeCu (30nm) bilayer with $\varphi=30°$ excited by a perturbation field at 2.86 GHz. (c) The distribution of $m_z$ along the $x$ direction at $t=20$ ns (red dotted line in (b)). (d) Dispersion relation of the FeCoSiB (30nm)/NiFeCu (30nm) bilayer with $\varphi=30°$.

The in-plane uniaxial anisotropic field $\vec{B}_{anis}$ is set as:

$$\vec{B}_{anis} = \mu_0 H_{ani}(\vec{u}\cdot\vec{m})\vec{u} \qquad (5)$$

with $\vec{u}$ representing the unity in-plane easy axis field vector, and the angle $\varphi$ can be changed by different $\vec{u}$.

Figure 1(b) plots the spatial and temporal distribution of $m_z$ for the FeCoSiB/NiFeCu bilayer with $\varphi=30°$ excited by a perturbation field at 2.86 GHz. The spatial distribution of $m_z$ along the red dotted line at 20 ns is shown in Fig. 1(c). Obviously, the wavelengths of the excited SWs along the $+k$ and $-k$ directions are significantly different, $\lambda=1.8\,\mu$m for $+k$ and $\lambda=2.6\,\mu$m for $-k$. In addition, the precession amplitude of SWs decreases with increasing propagation distance due to the damping of magnetic precession. The wavenumber values at the corresponding frequency in Fig. 1(d) can be extracted by Fourier transform of the time-domain results in Fig. 1(b). Therefore, using the perturbation field $\mathbf{h}$ mentioned before, the spin wave dispersion relations can be obtained, as shown in Fig. 1(d).

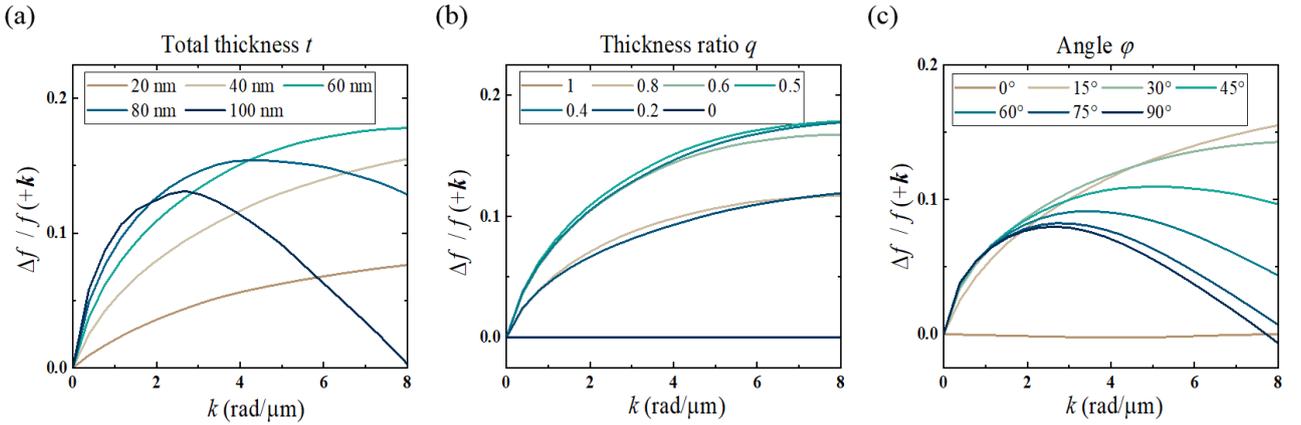

FIG. 2. Simulated $\Delta f / f(+k)$ of FeCoSiB/NiFeCu bilayers, (a) with different total thickness $t$, but fixed $q=0.5$ and $\varphi=15°$, (b) with different thickness ratio $q$ but fixed $t=60$ nm and $\varphi=15°$, and (c) with different angle $\varphi$ but fixed $t=40$ nm and $q=0.5$.

As shown in Fig. 1(c), the perturbation field can excite SWs with different wavelengths along the

+$k$ and −$k$ directions even for the same frequency. On the other hand, SWs of the same wavelength have different resonance frequency $f_{SW}$ in the ±$k$ direction. Thus, the normalized frequency shift ($\Delta f_{SW} / f_{SW}(+k)$) can be used to evaluate the magnitude of spin-wave nonreciprocity.

$$\Delta f_{SW} / f_{SW}(+k) = \frac{f_{SW}(-k) - f_{SW}(+k)}{f_{SW}(+k)} \qquad (6)$$

Figure 2 shows the SW nonreciprocity of magnetic bilayers with different total thickness $t$, thickness ratio $q$, and $\varphi$ angles. Here, $q$ is defined as the ratio of the NiFeCu layer thickness to the total thickness of the bilayer. As shown in Fig. 2(a), the increase of the total thickness is not only helpful to enhance the nonreciprocity, but also shifts the maximal $\Delta f_{SW} / f_{SW}(+k)$ to the low $k$ side. This can be attributed to the increasing contribution of interlayer dipole interaction [24]. Moreover, given a total thickness of 60 nm, there is no SWs nonreciprocity for single-layer FeCoSiB ($q = 0$) or NiFeCu ($q =1$) films, as shown in Fig. 2(b). For $q$ equals to 0.2 or 0.8, FeCoSiB/NiFeCu bilayers have similar $\Delta f_{SW} / f_{SW}(+k)$, and the same to $q$=0.4 or $q$=0.6. The strongest SWs nonreciprocity can be obtained at $q = 0.5$. As shown in Fig. 2(c), given a fixed $t$ of 40 nm and $q = 0.5$, $\Delta f_{SW} / f_{SW}(+k)$ quickly reaches its maximum at certain low wavenumber $k$, then its sign reverses at a higher $k$ at larger angle $\varphi$. This phenomenon is related to the energy of uniform and out-of-phase mode [24]. Although choosing a larger $\varphi$ angle seems beneficial to obtain greater SW nonreciprocity at low k, $\Delta f_{SW} / f_{SW}(+k)$ only differs slightly for $k \leq 2$. For 2<$k$≤8, the trend reverses, i.e., a low $\varphi$ angle is favored in order to obtain high nonreciprocity. In addition, $\Delta f_{SW} / f_{SW}(+k)$ is close to 0 at $\varphi = 0°$, again revealing that the SW nonreciprocity comes from the contribution of the interlayer dipole field, which vanishes at $\varphi = 0°$ [38].

**B. The coupling between SH-SAWs and nonreciprocal SWs**

In the previous part, the influences of different total bilayer thickness $t$, thickness ratio $q$ and angle $\varphi$ on the SW nonreciprocity are discussed, which provide preliminary guidance for the design of magnetoelastic bilayer structure. However, for magnetoacoustic devices, SWs need to be excited by SAWs via MEC. So, whether they can couple with each other at low wavenumbers and their coupling strength are also critical to obtain strong SAW nonreciprocity. Next, we will discuss the dispersion relations of interacting nonreciprocal SWs and SH-SAWs. It is recently reported that SH-SAWs are more efficient to excite SWs than well-studied Rayleigh-type SAWs [39]. We will limit our discussion within MEC since it is much stronger for magnetostrictive materials than other effects, such as spin-vorticity coupling [40] and magneto-rotation coupling [41].

The dispersion relation for the interacting SWs and SAWs has been studied in previous works [11,13,42], and was given as

$$\omega_{1,2} = \frac{\omega_{\mathrm{SW}} + \omega_{\mathrm{SAW}}}{2} \pm \sqrt{\left(\frac{\omega_{\mathrm{SW}} - \omega_{\mathrm{SAW}}}{2}\right)^2 + |\kappa|^2} \quad (7)$$

where $\omega_{\mathrm{SW}}$ and $\omega_{\mathrm{SAW}}$ represent the dispersion relations of the noninteracting SWs and SAWs respectively, and $\kappa$ is the coupling coefficient between SWs and SAWs. Although the phase velocity of SH-SAWs becomes dispersive upon coating a magnetoelastic layer [43], we here consider that $\omega_{\mathrm{SAW}}$ still satisfies the following linear relationship $\omega_{\mathrm{SAW}} = c_{\mathrm{SAW}} k$ due to small thickness of FeCoSiB layer, where $c_{\mathrm{SAW}}$ is the phase velocity of SH-SAWs (4165 m/s for LiTaO$_3$). Further, taking the damping rate and the SW nonreciprocity into account, Eq. (7) can be rewritten as

$$\omega_{1,2} - i\Gamma_{1,2} = \frac{(\omega_0 \pm \Delta\omega_{nr}) - i\Gamma_{SW} + \omega_{SAW} - i\Gamma_{SAW}}{2}$$

$$\pm \sqrt{\left[\frac{[(\omega_0 \pm \Delta\omega_{nr}) - i\Gamma_{SW}] - (\omega_{SAW} - i\Gamma_{SAW})}{2}\right]^2 + |\kappa|^2} \quad (8)$$

$\Gamma_{SW}$ and $\Gamma_{SAW}$ are the damping rate of SWs and SAWs, respectively. Due to the introduction of SW nonreciprocity, $\omega_{SW}$ is expanded to $\omega_0 \pm \Delta\omega_{nr}$, where $\omega_0/2\pi$ is the spin wave resonance frequency excluding the contribution from the interlayer dipole field [17]. $\Delta\omega_{nr}$ represents the strength of SW nonreciprocity of the bilayer structure, which is equal to $2\pi\Delta f_{SW}/2$ [14,17]. Its value is directly related to the space-broken interlayer dipole field, which has a complex relationship with the structural parameters and is not uniformly distributed in space [24]. For the reasons above, we do not attempt to write the analytical expression of $\Delta\omega_{nr}$, but will fit it using experimental results later.

For the SH-SAWs, $\kappa$ can be expressed as [11,39]:

$$\kappa = -\frac{2tb_1 \bar{\varepsilon}_{xy} m_{IP}^* \cos(2\varphi)}{\sqrt{A_k Q_k}} \quad (9)$$

where $b_1$, $\bar{\varepsilon}_{xy}$ and $m_{IP}^*$ represent the magnetoelastic coupling constant, the shear-horizontal strain and the in-plane dynamic component of magnetization averaged over the bilayer thickness, respectively. $A_k$ and $Q_k$ are the normalization constants [11]. Obviously, for SH-SAWs, $\kappa$ has the maximum value when $\varphi = 0°$ or $90°$, and vanishes for $\varphi$ close to $45°$. As illustrated in Fig. 3(a), for waves propagating in opposite directions, band gaps open at different frequencies and wave numbers, which correspond to the magnetoelastic coupling coefficient $\kappa$.

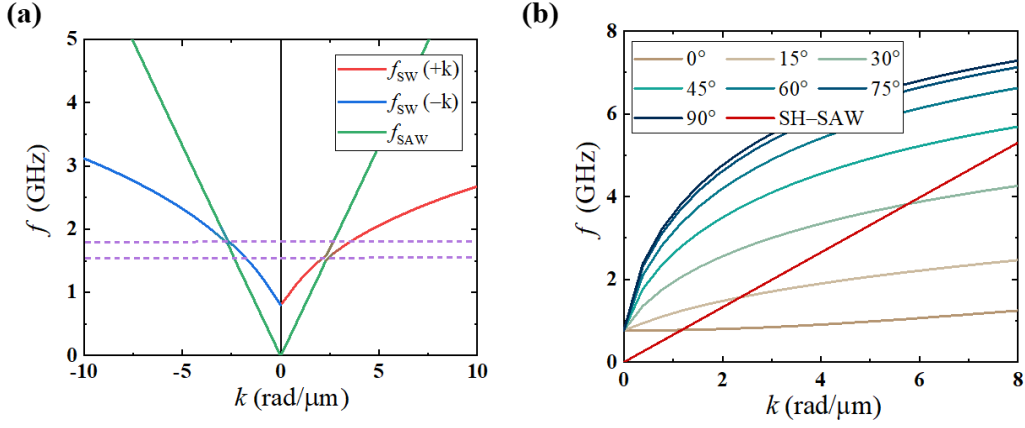

FIG. 3. (a) Dispersion relations of SH-SAWs and SWs of FeCoSiB (20nm)/NiFeCu (20nm) bilayer with $\varphi=15°$, where the purple dotted lines represent the coupling frequencies at $\pm k$ respectively. (b) Simulated SW dispersion relations along the forward $+\boldsymbol{k}$ direction with different $\varphi$ angles.

Figure 3(b) further plots the simulated dispersion relations of SWs of FeCoSiB (20nm) /NiFeCu (20nm) bilayer along the $+\boldsymbol{k}$ direction for different $\varphi$ angles. At the intersection points of SWs and SAWs, both $k$ and $f_{SW}$ increases rapidly with increasing $\varphi$ due to the increasing intralayer dipolar field [44]. Clearly, a relatively low $\varphi$ angle facilitates the coupling between SAWs and SWs at low $k$. In addition, considering the influence of $\varphi$ on the SW nonreciprocity of magnetoelastic bilayers in Fig. 2(c) and the angle dependence of coupling coefficient $\kappa$ in Eq. (9), a small $\varphi$ angle is necessary to obtain strong MEC as well as large SW nonreciprocity at low $k$.

Furthermore, although Fig. 2(a) shows that the increase of total thickness $t$ is helpful to enhance SWs nonreciprocity, larger mass load will lead to greater insertion loss. In addition, the increase of $t$ will also attenuate the effective strain $\bar{\varepsilon}_{xy}$, resulting in the decrease of the SWs-SAWs coupling coefficient. Therefore, thin bilayer structures will be more suitable for practical applications, as far as the interlayer dipole field is strong enough to support large SW nonreciprocity. This can be better facilitated by setting the thickness ratio $q$ as 0.5 according to Fig. 2(b).

## III. Experimental Methods

Single-layered $Fe_{70}Co_8Si_{12}B_{10}$, $Ni_{49}Fe_{21}Cu_{30}$ and bilayered FeCoSiB/NiFeCu films were prepared by magnetron sputtering on silicon substrates. An *in-situ* magnetic field was applied during sputtering to induce an in-plane uniaxial magnetic anisotropy (IPUMA). Figure 4(a) shows the *M-H* curves of FeCoSiB(10 nm)/ NiFeCu(10 nm) bilayer measured along the hard and easy axis by a vibrating sample magnetometer (VSM, BHV-525, Japan). Figure 4(b) shows the complex permeability spectra of FeCoSiB (20 nm) and NiFeCu (20 nm) as well as FeCoSiB (10 nm)/ NiFeCu (10 nm) bilayer using a well-established shorted microstrip transmission line perturbation method [45]. Owing to the low saturation magnetization of NiFeCu, the ferromagnetic resonance (FMR) frequency of the bilayer is significantly reduced compared with that of the single-layered FeCoSiB. The effective damping factor α≈0.018 was calculated by using $\Delta f(\mu'') = (\gamma/2\pi) M_s^{eff} \alpha$, where $\Delta f(\mu'')$ is the full width at half maximum (FWHM) of the imaginary permeability.

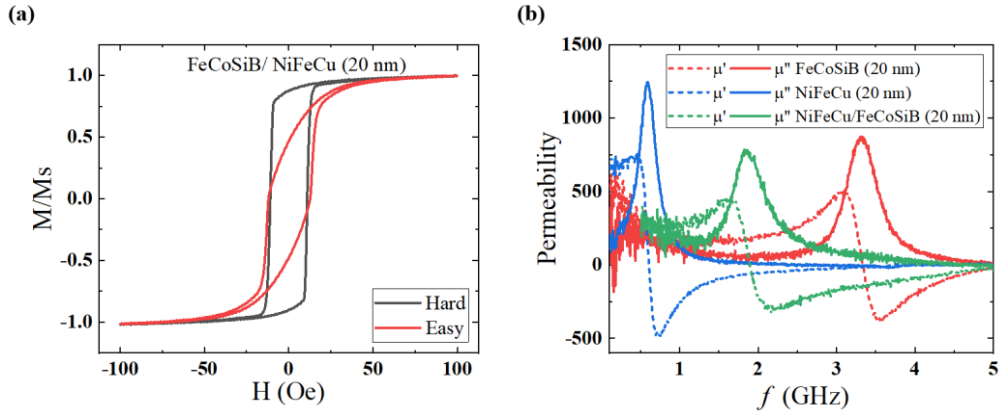

FIG. 4. (a) M-H curve for FeCoSiB (10 nm)/ NiFeCu (10 nm) bilayer. (b) Complex permeability spectra of single-layer FeCoSiB (20 nm) and NiFeCu (20 nm) as well as FeCoSiB (10 nm)/ NiFeCu (10 nm) bilayer.

We selected a 36°-rotated Y-cut X -propagation $LiTaO_3$ substrate to excite SH-SAWs. Aluminum split-finger interdigital transducers (IDTs) with a thickness of 50 nm were designed to reduce the reflection of SAWs and excited higher order harmonics. Delay lines with wavelength of 16 μm and spacing of 550 μm have been fabricated, which can excite SH-SAWs with fundamental frequency of 260 MHz and high order frequencies of 2.33 GHz ($SH_9$) and 2.86 GHz ($SH_{11}$). In the spacing between the IDTs, a 500×600 μm² rectangular FeCoSiB (10 nm)/ NiFeCu (10 nm) bilayer was deposited via sputtering and lithographically patterned. An *in-situ* magnetic field was again applied perpendicular to the SAWs propagation direction during sputtering to induce IPUMA. Figure 5(a) shows the photographs of a fabricated and packaged SAW device with FeCoSiB/NiFeCu bilayer. The two IDTs were connected to 50-Ω coplanar waveguides (CPW) on a PCB board via gold wires and the ends of the CPW were soldered to SMA connectors.

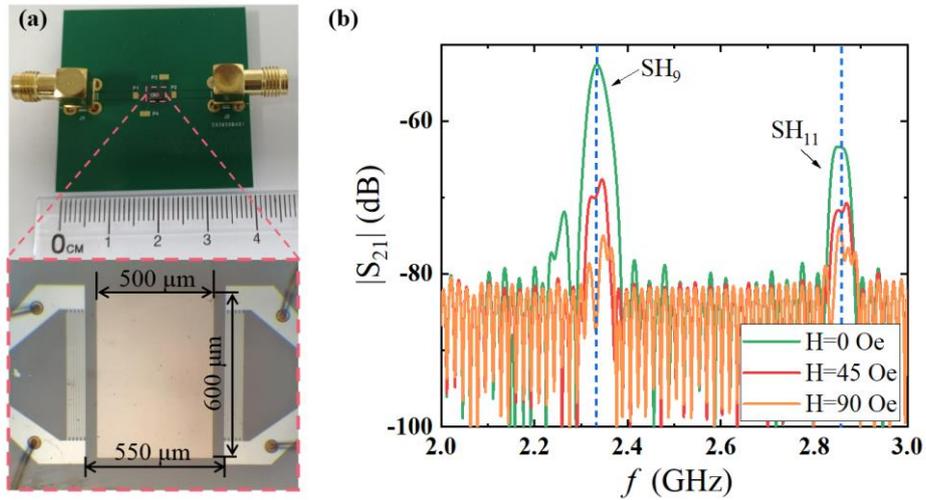

FIG. 5. (a) Photographs of a fabricated and packaged SAW delay line. (b) Measured $|S_{21}|$ of the delay line under different magnetic fields along $\varphi_H - \varphi_G = 0°$, and the blue dotted lines correspond to the frequencies of $SH_9$ (2.33 GHz) and $SH_{11}$ (2.86 GHz).

The transmission parameters of the delay lines were then measured by a vector network analyzer (VNA, Agilent N5230A) with an input power of 10 dBm. The device was placed in a Helmholtz coil powered by an ITECH-6502 DC source, and the output magnetic field was calibrated by a Gaussmeter (Lake Shore 425). The relative change of the background-corrected SAW transmission $\Delta S_{ij}(H)$ is

defined as

$$\Delta S_{ij}(H) = |S_{ij}(H)| - |S_{ij}(H_{225\,Oe})| \tag{10}$$

$|S_{ij}(H)|$ represents the transmitted parameters at acoustic resonance frequency $f_{SAW}$ under different external magnetic field **H** with $ij \in \{21, 12\}$, which is related with magnetoacoustic damping in Eq. (8):

$$|S_{21}| = -20\log(\Gamma_{1,2}L/c_{SAW}) \tag{11}$$

where $L$ represents the width of magnetic bilayer [15]. $|S_{ij}(H_{225\,Oe})|$ is the transmission parameters at a fixed field of 225 Oe, which is sufficient to saturate the bilayer. As seen in Fig. 5(b), $|S_{21}|$ changes upon applying magnetic field because of MEC, and $\Delta S_{ij}$ reflects the strength of the MEC coupling.

## IV. Results and Discussion

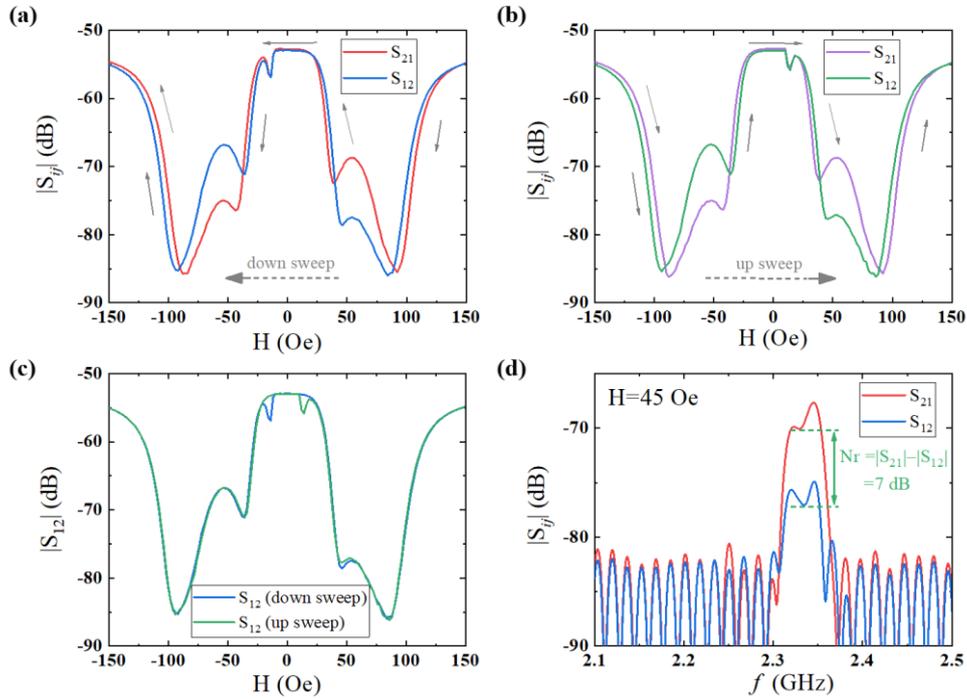

FIG. 6. (a)-(c) Comparison of $|S_{21}|$ and $|S_{12}|$ under the magnetic field of up sweep (-150 Oe to +150 Oe) and down sweep (+150 Oe to -150 Oe) at 2.33 GHz (SH$_9$). The external magnetic field is aligned at $\varphi_H - \varphi_G = 0°$. (d) $|S_{21}|$ and $|S_{12}|$ under $H = 45\,Oe$ along $\varphi_H - \varphi_G = 0°$ after time-domain gating [44].

Figs. 6(a) and 6(b) show the measured $|S_{ij}|$ under different magnetic fields during up sweep (-150 Oe to +150 Oe) and down sweep (+150 Oe to -150 Oe). The results during up and down sweeps are almost identical. The only difference is that the $|S_{ij}|$-H curves show a small dip near +15 Oe during up sweep, while they show a similar dip at -15 Oe during down sweep, as seen in Fig. 6(c). These dips are triggered by a reversal of magnetic moments near the coercive field [46]. It is worth mentioning that these small dips do not contribute to nonreciprocity, since the coercive field is far less than the field to induce strong MEC.

As mentioned in Sec. II, SWs propagating in $\pm k$ direction will couple with SAWs at different frequencies due to space-broken dipole field in the bilayer structure, leading to nonreciprocity [24]. Since SAWs can only be excited at specific harmonic frequencies for a given wavelength, this nonreciprocity is reflected in the fact that SAWs propagating in $\pm k$ direction reach the maximum $\Delta S_{ij}$ (the strongest MEC coupling) under different external magnetic fields. As shown in Figs. 6(a) and 6(b), the measured $|S_{21}|$ and $|S_{12}|$ separate from each other and then converge with the variation of magnetic field. Generally, the nonreciprocity (Nr) of SAWs transmission is defined as

$$Nr = |S_{21}(\mathrm{H})| - |S_{12}(\mathrm{H})| \tag{12}$$

Figure 6(d) plots the Nr at SH$_9$ under $H = 45\,\mathrm{Oe}$. Moreover, when the direction of the magnetic field reverses, one can find that the changes of $|S_{21}(\mathrm{H})|$ and $|S_{21}(-\mathrm{H})|$ are almost same, as seen in Figs. 6(a) and 6(b). This indicates that the reversal of either the external magnetic field or the SAWs propagation direction can lead to the appearance of nonreciprocity.

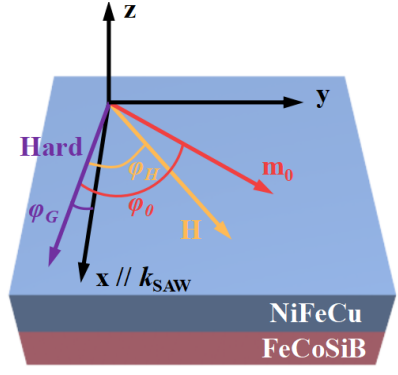

FIG. 7. The (*x*,*y*,*z*) coordinate system setting for measurement and calculation, which consists of the propagation direction of the SAW, the transverse in-plane direction, and the normal of the magnetoelastic bilayer.

Next, we measured the angular dependence $|S_{21}|$ upon applying external magnetic field. In order to facilitate discussion, we employed a (x,y,z) coordinate system, consisting of the propagation direction of the SAWs, the transverse in-plane direction, and the normal of the magnetoelastic bilayer. Here, $\varphi_G$ ($\varphi_H$) represents the angle between the *x* axis (the external magnetic field **H**) and the in-plane anisotropy hard axis. Thus, $\varphi_H - \varphi_G$ corresponds to the angle between the external magnetic field **H** and the SAW wave vector *k*, as shown in Fig. 7.

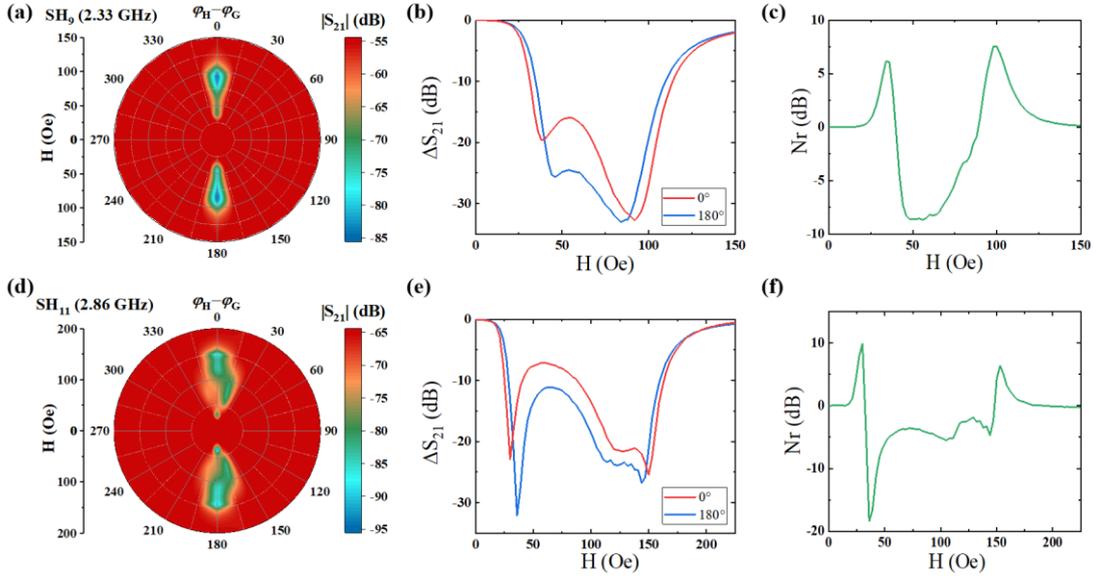

FIG. 8. (a) and (d) Polar plots of measured field-dependent $|S_{21}|$ of SH$_9$ (a) and SH$_{11}$ (d). (b) and (e) Measured $\Delta S_{21}$ along $\varphi_H - \varphi_G = 0°$ (or 180°) for SH$_9$ and SH$_{11}$, respectively. (c) and (f) Measured *Nr* along $\varphi_H - \varphi_G = 0°$ for SH$_9$ and SH$_{11}$, respectively.

Figures 8(a) and 8(d) plot the measured $|S_{21}|$ as a function of **H** and $\varphi_H - \varphi_G$ with a step of 15° for SH$_9$ and SH$_{11}$ modes, respectively. Both modes reach their maximum MEC at $\varphi_H - \varphi_G = 0°$ (or 180°), at which the magnetization gradually rotates towards the SAWs propagation direction ($\varphi$ close to 0°) upon increasing the external magnetic field. The SAWs-SWs coupling weakens and then disappears as $\varphi_H - \varphi_G$ shifts towards 90° (or 270°), at which $\varphi$ is always close to 90°. This angle dependence is consistent with Eq. (9), where the coupling coefficient $\kappa$ is proportional to $\cos(2\varphi)$. However, due to the angle dependence of the intralayer dipole field, the SWR frequencies increases significantly with $\varphi$ (see Fig. 3(b)). Therefore, although $\kappa$ can also reach its maximum at 90°, SH-SWs and SAWs cannot effectively couple because of the large difference between their resonance frequencies. What's more, the power absorption is strong at both 15° and 165° in Fig. 8(d), but becomes much weaker at 195° and 345°. This again reveals that the SW nonreciprocity ($\Delta\omega_{nr}$) has an angular dependence proportional to $\sin(\varphi)$, since the interlayer dipole field is proportional to $\sin(\varphi)$ [47].

In particular, the measured $\Delta S_{21}$ along $\varphi_H - \varphi_G = 0°$ and 180° are compared in Fig. 8(b) and 8(e) for SH$_9$ and SH$_{11}$ modes, respectively. For both modes, about -32 dB of power absorption can be observed, indicating strong SWs-SAWs coupling. When the magnetic field direction changes from 0° to 180°, the field corresponding to the minimum $\Delta S_{21}$ shifts. As shown in Fig. 8(c) and 8(f), the maximum nonreciprocity of the SH$_9$ mode is 8.6 dB at $H = 50\,\text{Oe}$, while that of SH$_{11}$ mode increases to 18.6 dB at $H = 36\,\text{Oe}$. The enhancement of nonreciprocity can be attributed the increase of wavenumber *k*, which is consistent with the trend in Fig. 2. It should be noted that $\varphi$ is not 0° under external magnetic fields of 25-150 Oe even for $\varphi_H - \varphi_G = 0°$, otherwise the nonreciprocity should be zero. This may be explained by (i) deviation of $\varphi_G$ from 0° during film deposition, and (ii) the angular dispersion of magnetic moments arising from stray fields and local stress/structural inhomogeneities.

According to the measured $\Delta S_{21}$ in Figs. 8(b) and 8(e), we can derive the maximum coupling coefficient $\kappa$ by solving Eqs. (8) and (11), $2\pi \times 76$ MHz for SH$_9$ and $2\pi \times 74$ MHz for SH$_{11}$. Here, $\Gamma_{SAW} = 2\pi \times 1.5$ MHz and $\Gamma_{SW} = \alpha\varepsilon\omega_{SW}$ were used for calculation following Ref. [48] with the ellipticity-related factor $\varepsilon$ of 3 and measured $\alpha$ of 0.018. The maximum magnetoacoustic damping can be calculated as $\Gamma_{1,2} = (\Gamma_{SW} + \Gamma_{SAW})/2$ for $\kappa > (\Gamma_{SW} - \Gamma_{SAW})/2$, $\sim 2\pi \times 63.7$ MHz, consistent with that directly obtained from Eq. (11). Additionally, the term of $\omega_0$ in Eq. (8) can be calculated by using Eq. (13) developed in our previous work [44].

$$\omega_0 \approx \gamma\sqrt{\frac{\left(H\cos(\varphi_0 - \varphi_H) - H_{ani}\cos(2\varphi_0) + M_s^{eff}(\frac{|k|t}{2})\sin^2(\varphi_0 - \varphi_G)\right) \cdot \left(H\cos(\varphi_0 - \varphi_H) - H_{ani}\cos^2\varphi_0 + M_s^{eff}\right)}{1 + \alpha^2}} \quad (13)$$

where $M_s^{eff}$ represents the effective saturation magnetization for the magnetic bilayer [27,28]:

$$M_s^{eff} = qM_s^B + (1-q)M_s^A \quad (14)$$

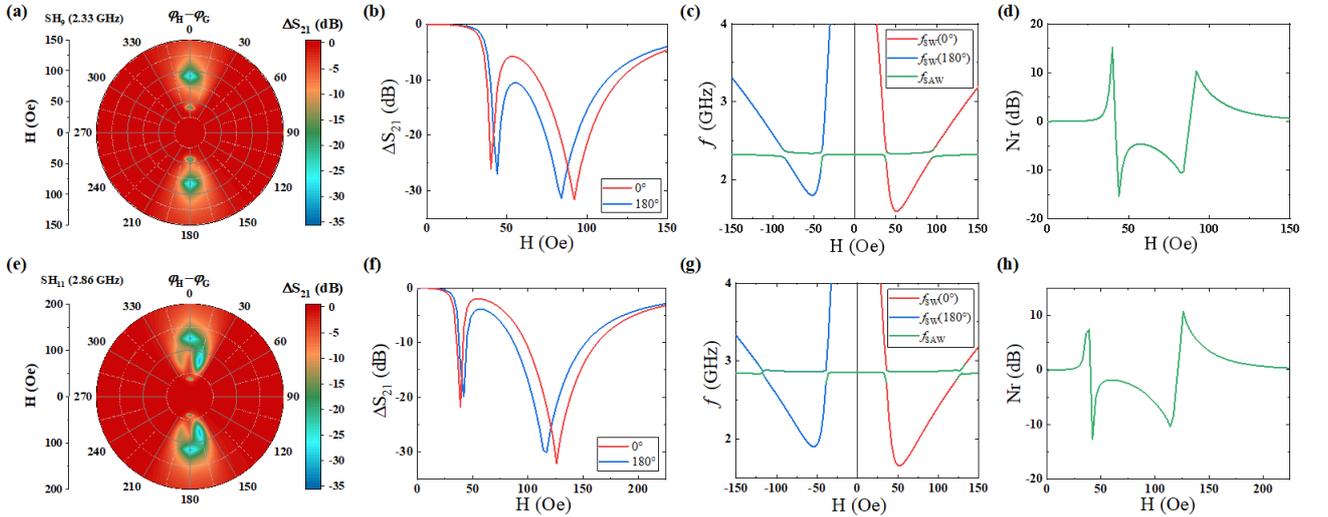

FIG. 9. (a) and (e) Polar plots of calculated field-dependent $\Delta S_{21}$ of SH$_9$ and SH$_{11}$. (b) and (f) Calculated field dependent $\Delta S_{21}$ along $\varphi_H - \varphi_G = 0°$ (or 180°) for SH$_9$ and SH$_{11}$, respectively. (c) and (g) Calculated field dependent $f_{SW}$ along $\varphi_H - \varphi_G = 0°$ (or 180°) coupled with SH$_9$ and SH$_{11}$. (d) and (h) Calculated field dependent $Nr$ along $\varphi_H - \varphi_G = 0°$ for SH$_9$ and SH$_{11}$, respectively.

The polar plots of fitted field-dependent $\Delta S_{21}$ of SH$_9$ and SH$_{11}$ are given in Figs. 9(a) and 9(e) with the in-plane uniaxial anisotropic field $H_{ani} = 35$ Oe and $\varphi_G = 3°$. It can be seen that the experimental results are in good agreement with the fitting ones in terms of both angle and magnitude dependencies. Notice that our fitting results indicate that the IPUMA slightly deviates from the x-axis, which explains why $\varphi$ is not 0° under external magnetic fields for $\varphi_H - \varphi_G = 0°$. Moreover, according to the experimental results, $\Delta\omega_{nr}$ can be further rewritten to $\Delta\omega_{nr}'\sin(\varphi)$ by taking the $\varphi$ angle dependence into consideration [47]. And the fitted $\Delta\omega_{nr}'$ values are 2π×491 MHz for SH$_9$ and 2π×598 MHz for SH$_{11}$. What's more, the $\Delta S_{21}$-$H$ curve along $\varphi_H - \varphi_G = 0°$ (or 180°) show two dips upon varying magnetic fields in Figs. 8(b) and 8(e) as well as Figs. 9(b) and 9(f). When the external magnetic field is applied along the hard-axis direction of the magnetic bilayer, $f_{SW}$ will first decrease and then increase. Thus, SH-SAWs and SWs can couple at two magnetic fields, as shown in Figs. 9(c) and 9(g). For comparison, the calculated $Nr$ for SH$_9$ and SH$_{11}$ also show in Figs. 9(d) and 9(h), which are again close to the measured results in Figs. 8(c) and 8(f).

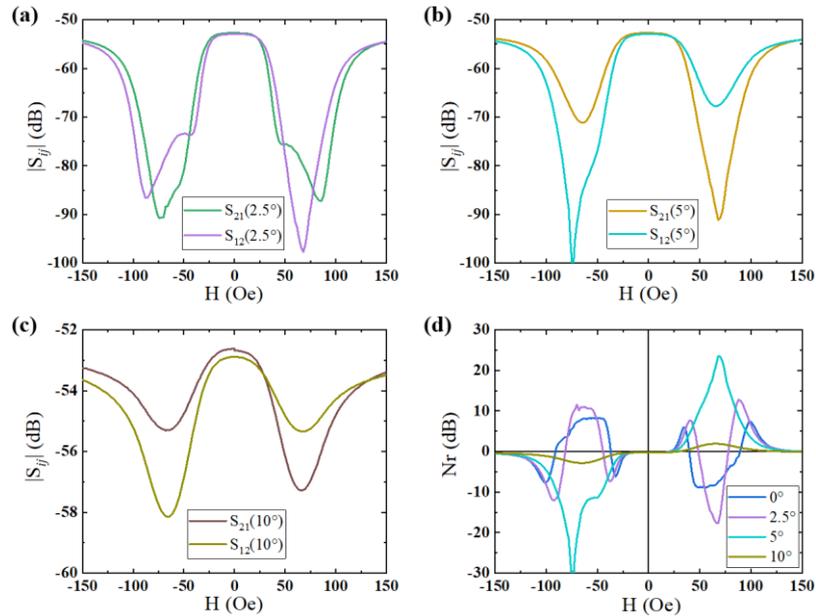

FIG. 10. (a)-(c) Measured $|S_{ij}|$ under different magnetic fields at 2.33 GHz (SH$_9$) along various angles $\varphi_H - \varphi_G$. (d) Comparison of measured $Nr$ under different magnetic fields at various angles $\varphi_H - \varphi_G$.

Finally, in order to further clarify the angle dependence of nonreciprocity, $|S_{ij}|$ was measured at 2.33 GHz (SH$_9$) upon varying $\varphi_H - \varphi_G$ in a small range between 0 and 10°, as shown in Figs. 10(a)-10(c). As $\varphi_H - \varphi_G$ increases, Nr first increases and then decreases (Fig. 10(d)), and the largest nonreciprocity about 30 dB (or 60 dB/mm) is achieved at $\varphi_H - \varphi_G = 5°$. This is kind of odd, since the interlayer dipole field of FeCoSiB/NiFeCu bilayer is proportional to $\sin(\varphi)$ as discussed above. Thus, given the same external magnetic field, both $\varphi$ and the interlayer dipole field become larger as $\varphi_H - \varphi_G$ increases. However, one should notice that the SWR frequency is also closely related to $\varphi_H - \varphi_G$. Figure 11 plots $f_{SW}$ as a function of the external magnetic field for $\varphi_H - \varphi_G = 0°, 2.5°, 5°$ and 10°. Clearly, the $f_{SW}$-H curve can intersect with $f_{SAW}$=2.33 GHz at two points for $\varphi_H - \varphi_G = 0°$ and 2.5° along both positive and negative field directions, but only insects with it at one point or even not for $\varphi_H - \varphi_G = 5°$ and 10°. This not only explains the double and single $|S_{ij}|$-H curve dips in Figs. 10(a)-(c), but also highlights the importance of MEC coupling discussed in Sec. II B. The measured nonreciprocity at $\varphi_H - \varphi_G = 10°$ is significantly reduced due to the weak MEC coupling between SAWs and SWs.

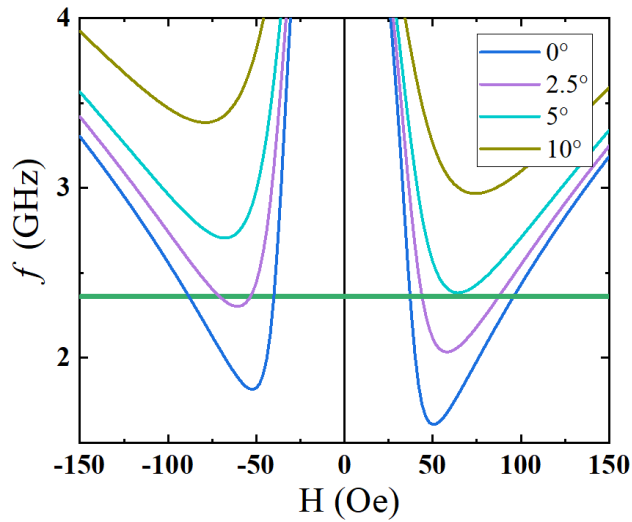

FIG. 11. Calculated $f_{SW}$ under different magnetic fields along various $\varphi_H - \varphi_G$ angles. The green line corresponds to the center frequency of SH$_9$ at 2.33 GHz.

## V. Conclusion

In summary, we demonstrated large nonreciprocity in SH-SAW delay lines coated with a magnetoelastic bilayer. Although static *M-H* and high frequency permeability spectrum measurements indicate that the two layers are strongly exchanged coupled, the dipole interaction can be excited by SAWs with certain wavenumber, thus breaking the space symmetry in the thickness direction. The designed large difference in the saturation magnetization of FeCoSiB and NiFeCu not only promotes strong interlayer dipole interaction, but also ensures the SAWs-SWs coupling at relatively low wavenumbers. The impacts of geometric and magnetic parameters on the SW nonreciprocity have been studied by means of micromagnetic simulations. Of particular importance is the $\varphi$ angle between the magnetization and the SAW propagation direction, which affects both interlayer and intralayer dipole interaction. The dispersion relation of interacting SH-SAWs and nonreciprocal SWs was subsequently developed in consideration of magnetoelastic coupling. These efforts lead to a large nonreciprocal transmission of SH-SAWs about 30 dB (i.e., 60 dB/mm) at a low wavenumber of 3.53 rad/μm (2.33 GHz). The nonreciprocity could be further improved for SH-SAWs devices with much higher frequency, which can meet the conditions for both MEC coupling and interlayer dipole interaction at $\varphi = 90°$. One obvious concern to address is the high forward insertion loss, which can be mitigated by employing solid mounted piezoelectric thin films and advanced IDT designs [49,50].

**Acknowledgement**

This work is supported by the National Natural Science Foundation of China (Grant No. 61871081 and 61271031) and the Natural Science Foundation of Sichuan Province under Grant No. 2022NSFSC0040.

MEMS, 1030-1033 (2022).

[50] L. Zhang, S. Zhang, H. Zhou, J. Wu, P. Zheng, H. Xu, Z. An, T. You, and X. Ou, High Frequency, Low Loss and Low TCF Acoustic Devices on LiTaO3-on-SiC Substrate, 2021 IEEE IUS, 1-4 (2021).
26